# Energy harvesters for space

R. Graczyk

*Abstract*—Energy harvesting is the fundamental activity in almost all forms of space exploration known to humans. So far, in most cases, it is not feasible to bring, along with probe, spacecraft or rover suitable supply of fuel to cover all mission's energy needs. Some concepts of mining or other form of manufacturing of fuel on exploration site have been made, but they still base either on some facility that needs one mean of energy to generate fuel, perhaps of better quality, or needs some initial energy to be harvested to start a self sustaining cycle. Following paper summarizes key factors that determine satellite energy needs, and explains some either most widespread or most interesting examples of energy harvesting devices, present in Earth orbit as well as exploring Solar System and, very recently, beyond. Some of presented energy harvester are suitable for very large (in terms of size and functionality) probes, other fit and scale easily on satellites ranging from dozens of tons to few grams.

*Index Terms*—energy harvesters, orbital battery, radioisotope, satellite, Seeback effect, solar array, tether, thermoelectric generator,

## I. Introduction

Energy harvesting is a concept of gathering energy from ambient, often using renewable energy source. Harvested energies are of various kind, including one transmitted in electromagnetic waves of different lengths – from classic radio frequencies used for communication to sunlight, thermal gradients, kinetic and potential energies. Generally speaking, sources of energies that are being gathered are of two types. First is the intrinsic energy located in environment where harvester operates (like Sun providing power to photovoltaic cells). Second is the waste energy dissipated by other operating devices (like electric motor providing power to thermal or kinetic harvester). Energy, when collected, usually is converted into electricity and stored. Power generated by portable, autonomous harvester devices, depending on technology used spans from ten of microwatts to tens of miliwatts. This fact implies that systems powered by energy harvesting device have to be small in terms of power needs. As a result, functionality of such systems has also to be limited, especially that they are likely to operate in charge (energy build-up in span of minutes or tens of minutes) and act (perform the measurement and processing quickly and effectively).

State-of-the-art research facilitated by growing industrial interest in energy harvesting as well as innovative products introduced into market recently show that described techniques are viable source of energy for sensors, sensor networks and personal devices. Research and development activities include technological advances for automotive industry (tire pressure monitors, using piezoelectric effect ), aerospace industry (fuselage sensor network in aircraft, using thermoelectric effect), manufacturing automation industry (various sensors networks, using thermoelectric, piezoelectric, electrostatic, photovoltaic effects), personal devices (wrist watches, vital signs monitors, remote controls, using piezoelectric and thermoelectric effects).

Activities and current research in mentioned field focus on devices that are miniaturized, very low power, harvest energy on low scale (which is the only suitable for them). Those solutions aim to guarantee sensor autonomy (no or wireless communication), safety (no power lines) and no or minimal maintenance.

It is worth noting that those design drivers which are listed above, are almost exactly the same as baseline requirements for all human made artificial Earth satellites and deep space probes exploring Solar System and its boundaries (and, very recently, beyond [reference to Voyager]). Almost every satellite and probe is an energy harvesting device, performing energy harvesting on much greater scale as it is done in examples shown before. The reason for that fact is simple. Satellites have to operate continuously and it is currently not possible, from both, technological (tricky rendez-vous maneuver, reserved only for largest parts of space infrastructure ) and economical (cost of one kilogram of mass sent to Low Earth Orbit, LEO varies from 10k to 100k euro) to resupply any of the fuel like consumable that could be used on board.

Satellites and probes vary significantly in terms of their size, weight and applications. Satellites energy needs, technology used for energy collection, and the source of harvested energy varies as well. On average, satellite (or probe, which are used interchangeably in this paper) size can be from several meters and several tons (large satellites, typically telecommunication, military earth observations, telescopes) to few centimeters and few hundred grams (micro-

Rafał Graczyk is a PhD student in field of high reliability electronics and FPGA applications at Electronics and Information Technology Faculty at Warsaw University of technology and works as electronics specialist in Space Research Centre of Polish Academy of Sciences. He is leading a team responsible for development and testing of On Board Computers and Electrical Power Subsystem for first Polish scientific satellites BRITE-PL "Lem" and "Hevelius" (e-mail: rgraczyk@cbk.waw.pl).

to nano-satellites, technology experiments, scientific and amateur radio). Similarly power needs of satellites varies from several kilowatts for TV broadcasting relays to hundreds of miliwatts for satellites-on-chip experimental designs. Sources of energy for powering the probe are different, starting from most popular photovoltaic solar cell arrays present in almost all satellites operating in inner Solar System. Second popular, the thermoelectric generators, working on temperature difference between two heat storages, heated by solar infrared radiation (or radioisotopes) and cooled by cosmic background. Fuel cells and nuclear reactors are also examples of energy sources for satellites but they are definitely not main stream. Fuel cells were used in Space Shuttle, which had a limited operation time capability and now is withdrawn from use. Nuclear reactors pose a serious threat to environment in (not so unlikely) case of launch failure, and due to is heavy infrared signature (very easy to detect) are not used even for military purposes. There is also another energy source to be used by spacecraft, recently emerging as a very interesting implementation option, especially for micro- and smaller satellites – electrodynamic tethers.

Out of the energy sources in use in orbit all base on harvesting energy forms which are widely present there – solar radiation (in visible and IR ranges) and a gravity combined with Earth's magnetic and plasma environment. Following chapters introduce the basic principles of operation of photovoltaic, thermoelectric and tethered harvesters used or intended for use on board of probes, rovers and satellites [10].

## II. Photovoltaic effect harvester

Earth orbiting spacecraft, ranging from low-Earth orbit to geosynchronous orbit, typically employ solar photovoltaic (PV) array. Rationale, for PV array implementation over vast and diverse classes of satellites is simple: solar PV array are easily scalable from 100W, even up to 300kW, have decent specific cost ($/W), offer unlimited access to "fuel" and have minimal safety analysis reporting needs.

Solar arrays are assembled from large number of individual solar cells arranged on a mechanical supporting panel which convert solar energy into electric power by photovoltaic conversion. The solar cells are made in various shapes and sizes which put out relatively low current and voltage. The first application of solar photovoltaic power dates back to March 17, 1958 when Vanguard 1 was launched utilizing six solar panels which provided less than one watt of power for over six years with a 10% conversion efficiency. Solar cell design is rated by its ability to convert a certain percentage of the solar energy into electric power which is known as the solar cell efficiency which is defined as Eq. 1

$$\eta = P_{out} / P_{in} \qquad (1)$$

where $P_{out}$ is the electrical power output and $P_{in}$ is the solar power input.

Solar cells are organized to cover as much area as possible, to exhibit as much of insolation as possible. The solar constant, equal to 1358 W/m$^2$, is the total solar energy incident on a unit area perpendicular to the sun's rays at the mean Earth-Sun distance outside the Earth's atmosphere. It varies between about 1310 and 1400 on an annual cycle, with maximum at perihelion and minimum at aphelion. Solar cell can take most of this power when are set at zero degrees in respect to incoming rays. Raising this angle lowers the effective irradiation power trigonometrically. That's a reason for implementing sophisticate attitude determination and control systems on board of larger spacecraft, to ensure solar array takes most out of incoming light with respect to current orbital operations (satellites mission).

Solar cells are connected in series to maximize voltage and in parallel for current. To minimize power losses with a single cell failure, the solar array cells are connected in a series parallel ladder network.

It's important to note that solar array must collect enough energy during sunlight to power spacecraft during entire orbit (eclipse time can take around 40-45% of whole orbit period).

The cell design is affected by various factors which must be considered such as I-V characteristics, its temperature dependence, distance from sun, and radiation degradation. The current-voltage (I-V) characteristics of solar cells are of great importance in the design of solar arrays. PV array is aimed to be designed for minimum mass and maximum efficiency at the maximum power point (MPP). The MPP is where the product of I and V is at a maximum which is defined by the maximum area rectangle within the plot. This point always falls at the knee of the I-V curve (Fig. 1) [10].

There are two methods used for working with maximum power point of an array. First one, typically used for smaller satellites is to use Direct Energy Transfer control. Direct Energy Transfer (DET) systems dissipates unneeded power Typically use shunt resistors to maintain bus voltage at a predetermined level  Shunt resistors are usually at the array or external banks of resistors to avoid internal heating. It means that solar array has to be sized such to cover worst case insolation case and worst case satellite power needs (and to recharge batteries for eclipse).

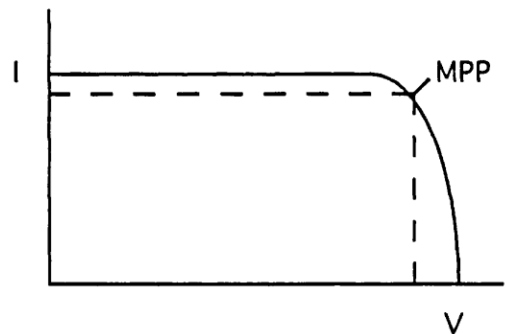

Fig. 1 Photovoltaic cell current-voltage curve

Other method, more sophisticated is to implement Peak Power Tracker distribution system. Peak Power Trackers (PPT) extract the exact power required from the solar array. Some DC/DC converters have to be put in series with the array in order to dynamically change the solar array's operating point  - according to changing system needs. Typically it requires 4 - 7% of the solar array power to operate, and are often used on larger satellites [8].

### III. THERMAL ENERGY HARVESTER

There are two types of harvesters transforming heat into electricity used in space applications. First of them is static thermal energy harvester, typically a thermoelectric generator built around radioisotope source. The second type of power generator subsystem, often applied in GPS constellation satellites is  AMTEC alkaline electrochemical cell, which is an example of dynamic thermal energy harvester.

The physical phenomena that serves as a base for creation of thermoelectric harvester is Seeback effect. Seeback effect is a process of direct heat gradient into electricity conversion over a two different materials junction. If it is assumed that one material is in temperature $T$ and the other material forming junction is in temperature $T + \Delta T$, then the resulting junction voltage $\Delta V$ is proportional to difference $\Delta T$. The proportion $\Delta V / \Delta T$ is characteristic to material used and is called Seeback coefficient (thermopower) (Eq. 2).

$$\alpha \approx \Delta V / \Delta T \qquad (2)$$

The Seeback coefficient value may vary significantly and ranges from ~1mV/K (for metals) to ~100mV/K (for semiconductors). Seeback effect has also a reversed counterpart called Peltier effect, also widely used in space application – for refrigeration purposes, as cooling of some electronic part is heavily limited due to vacuum and lack of convection (which constitutes for 80% of heat transfer when it is taking place in terrestrial conditions). Similarly, Seeback effect is used in all thermoelectric generators that have been customized for operation in outer space.

Simple model of Peltier cell (left) and Seeback cell (right) built on semiconductor chunks are shown on Fig 2. When heat is absorbed on top side of a thermoelectric generator the movable charge carriers begin to diffuse, creating an uniform concentration distribution in the device along the temperature gradient, and producing the difference in the electrical potential on both sides of it. As it has been explained, due to Seeback effect, electrons flow through the n-type semiconductor chunk to the colder side while in the p-type semiconductor, the holes flow to the cold side. To maximize the power generation output, thermoelectric cells are connected together in chains (thermally in parallel and electrically in series) to harvest as much heat as possible [13,14].

Most of the space applications of thermoelectric generators assume coupling with some radioisotope heat source. In fact, NASA designed such standard unit called Radioisotope Thermoelectric Generators (RTGs). Up to the time of the writing of this paper 41 RTGs has been launched on 23 spacecraft. Their power production ranges from 2 watts to 300 watts. Total system efficiency is around 7% (~2kW heat production in order to obtain 130 W of usable spacecraft power). The RTGs use Plutonium-238 with an 87.7-year half-life [10].

To provide higher efficiencies of electric power production, the development of space solar dynamic power systems has been proposed. The difference between thermal static and thermal dynamic power is the power conversion technique. Instead of direct conversion of heat into electricity as with thermoelectric effect, solar dynamic systems use solar power to heat a working fluid to drive a heat engine which is used to generate electricity. The advantage of solar dynamic systems over solar photovoltaic systems is that dynamic systems in general have a higher thermal efficiency and can be used for higher power levels. A solar dynamic system consists of four basic components, the collector/concentrator, receiver, radiator, thermal storage material, and the heat engine turning heat into movement. Movement is then turned into electricity in alternator. The power conversion cycle can be any of the common thermodynamic cycles: Rankine, Brayton, or Stirling. Similar thermal dynamic power generator, but working on slightly different principle, is a electrochemical concentration cell that converts the work generated by thermal expansion of sodium vapor directly into electric power, called AMTEC.

Solar AMTEC (alkali metal thermal-to electric conversion) is a new power system concept implemented recently in new generation of Global Positioning System satellites. SAMTEC system consists of array of multi-tube, vapor anode AMTEC cells (having 24% conversion efficiency) and a direct solar irradiation receiver integrated with LiF salt canisters performing a role of energy storage even during maximum eclipse periods. Due to effective heat energy storages electrical battery packs storage capacity is much less necessary. AMTEC cells are connected in series and parallel and each cell is redundant in order to achieve seamless 10-12 years in orbit operation without reduction of energy conversion efficiency due to background radiation, which would be serious issue in case of PV arrays. AMTEC cell which is the heart of described harvester. The device accepts a heat input in a range from about 900 K – 1300 K and produces direct current with expected efficiency between 15-40%. In the AMTEC sodium is flowing around a closed thermodynamic cycle between a high temperature heat reservoir (salt canister) and a cooler reservoir at the heat rejection temperature (radiator fins). The unique thing about AMTEC cycle is that sodium ion conduction between a high pressure or activity region and a low pressure or activity region on either side of a ion-conducting refractory solid electrolyte, is thermodynamically nearly equivalent to an isothermal expansion of sodium vapor between the same high and low pressure. Electrochemical oxidation of neutral sodium at the anode leads to sodium ions which traverse the solid electrolyte and electrons which travel from the anode through an external circuit where they perform electrical work, to the low pressure cathode, where they recombine with the ions to produce low pressure sodium gas. The sodium gas generated at the cathode then travels to a condenser at the heat rejection temperature of perhaps 400–700 K where liquid sodium reforms. The AMTEC thus is an electrochemical concentration cell which converts the work generated by expansion of sodium vapor directly into electric power [3].

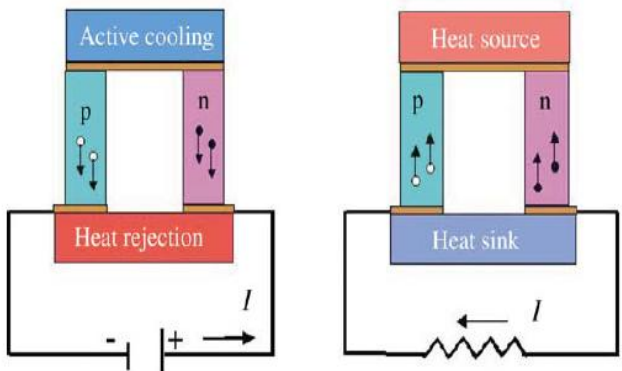

Fig. 2 Semiconductor thermocouples depicting Seeback and Peltier effects.

Typical SAMTEC Electrical Power Supply configuration consists of a symmetric, rigid parabolic concentrator supported by rods extruding from front face of solar receiver. Typical interior of receiver assembly. Each receiver consists of AMTEC cells and TES canisters, short sodium heat pipe and a heat receiver plate (called hot shoe). Cells and canister assemblies are arranged in a way to form the solar receiver cavity walls. Solar radiation enters through aperture and falls mainly on sidewalls, and heats canister and side of AMTEC cells. Heat receiver plates are designed in such way they absorb incident solar energy and radiate part of it to lower parts of solar receiver cavity. Thermal gradient is formed along main axis. Such device provides around 1200 W of power available to spacecraft bus [11,12].

## IV. ELECTRODYNAMIC TETHER HARVESTER

A space tether is a long thin structure (line like) that extends from a sub-orbital or orbiting spacecraft such as a rocket, satellite, or space station. Space tether length can vary from single meters up to 30 or so kilometers (world record is bit more than 32 kilometers). Tethers have to be of extraordinary strength to withstand tensile forces. For tethers, three categories of applications could be distinct: momentum exchange, formation flying and electrodynamic. Mechanical tethers (first two categories) connect masses on orbit but doesn't have to be conductive. Electrodynamic tethers (EDTs) are conductive. EDT can be built using any conductive material, and the tradeoff between the mass and resistivity is matter of case by case analysis (typically it is aluminum or copper).

EDT operation principles are straightforward. First, as the tether moves along its orbital path, there is an electromotive force generated along it. Second, tether wire is a low resistance path connecting regions of different plasma density and parameters. Third, connection to ionosphere can be limited only to tether ends, or can be continuous along tether length which affect the electrodynamic properties of whole system.

Electromotive force is a Lorentz force that results from tether (long conductor) movement in Earth's electromagnetic field (Eq. 3).

$$F = q(E + v \times B) \quad (3)$$

where $q$ is the charge of an electron, $E$ represents any ambient electric field (small), and $v \times B$ represents the motional electric field as the tether travels at a velocity $v$ through the Earth's magnetic field, represented by $B$. Equation 3 can be rewritten as

$$F = qE_{tot} \quad (4)$$

where :

$$E_{tot} = E + v \times B \quad (5)$$

and represents the total electric field along the tether.

In order to get the total electromotive force generated along the tether, $E_{tot}$ must be integrated along the entire length of the tether. Resulting total tether (of length $l$) potential is

$$\varphi_{tether} = -\int_0^l E_{tot}\, dl \approx -\int_0^l v(l) \times B(l)\, dl \quad (6)$$

which is negative as electrons in the tether are acted upon by the Lorentz force. As the ionosphere plasma surrounding the EDT decent conductor and ambient electrostatic field of $E$ is small and can be ignored. The tether potential is path independent assuming a conservative resultant electric field and steady-state conditions. As a result value of $\varphi_{tether}$ depends only on relative locations of the wires endpoints (their separation distance and orientation) and does not depend on the position of the tether between both ends.

The second and third principles, mentioned at the beginning of a chapter are related to current flow through the tether, which occurs when a connection is made between wire ends and orbital plasma. This can be done passively or enforced actively. Passive solution requires voltages in the tether –plasma system distribute themselves in a self-consistent manner, which can require very high level of tether charging in order to enable current flow [1,2].

Active solution requires electron generator device, such as an electron emitter or hollow cathode. Both solutions cause current flows through the wires as in attached picture (Fig. 1) In the passive solution, current flows up the tether because the resultant force on the electrons is downwards (towards Earth). After electrons are collected at the remote end of tether, where the counter mass, a sub-satellite is located, they are conducted through the tether to main spacecraft where they are ejected into surrounding plasma. Current closure occurs in the ionosphere, completing the electrical circuit.

If current flows in the tether element, a force is generated and can be expressed by

$$F = \int_0^l I(l) \times B(l)\, dl \quad (7)$$

When tether is in energy-harvesting (or, in other words, de-orbit mode), the electromotive force can be used by the tether system to supply the current from the tether into electrical loads, including energy storing devices like flywheels and batteries), eject electrons at the emitting end or grab electrons at the counterbalance end.. In boost mode (orbit raising mode), on-board power supplies must overcome electromotive force resulting from travel through magnetosphere, to drive current in the opposite direction, therefore creating a force in the opposite direction than Earth, boosting up the system. Thrust levels vary significantly on applied power and typically are less than 1 N, which is still three to six orders of magnitude better than these generated by other electric propulsion like ion or Hall-effect engines.

Numeric simulations, verified by experimental results, show that for large satellite a 10 km long electrodynamic tether can provide average power of 1kW (2kW in peak) with 70-80% efficiency of potential to electric energy conversion. Typically such tether, if made out of aluminum, weights almost 8kg which is almost nothing when compared to thousand or more kg for a large Earth orbiting satellite [4,5].

Smaller satellite platforms also are considered as hosts for electrodynamic tether, namely nanosatellites called CubeSats. Cubsats, depending on configuration, typically does not exceed 4 kg and 30 cm in largest dimension for 3 unit model. An experimental prototype has been built, allocation 1.33 kg – 1 unit for all EDT components, including end-mass, spool , deployment controller and an electron gun. For tether, low mass variant of AWG 25 aluminum wire has been selected. Length of 1300 m (resulting in tether mass bit below 0.6 kg) is a breakeven point, where tethered system can generate the same amount of power as state-of-the-art photovoltaic array of size that could be reasonably accommodated on 3 unit Cubesat.

The extraordinary growth of nanosatellite applications has facilitated increase in interest of even smaller spacecraft, pico-satellites, satellites-on-chip operating in distributed swarm. Advances in electronics design and micro-electro-mechanical systems paving their way on board of pico-satellites have proven feasibility of construction and applicative usability. Problem, although, lies in orbital dynamics. Satellites on chip suffers from low ballistic coefficients. They have very low mass (thus, kinetic energy) while having relatively high area of average cross-section (thus, aerodynamic drag). As a result, their orbital life-time ranges from days to hours. As a solution, to counter the drag force, a short, semi-rigid electrodynamic tether for propulsion is considered, which keeps the overall satellite-on-chip mass low and efficiently provides enough thrust to overcome drag in low Earth orbit. In year 2000, DARPA performed an experimental flight of two pico-satellites (100 g, 5 cm × 5 cm × 2 cm) connected with 30 m long electrodynamic tether. In course of mission, satellite-tether-satellite system performed scheduled operations, including boosting orbit altitude, energy harvesting and some more sophisticated formation flight maneuvers. All maneuvers were satisfactory to an extent which led the scientist to consider tethered femto-satellite swarms (1g, 1cm x 1 cm 1 cm) [6,7].

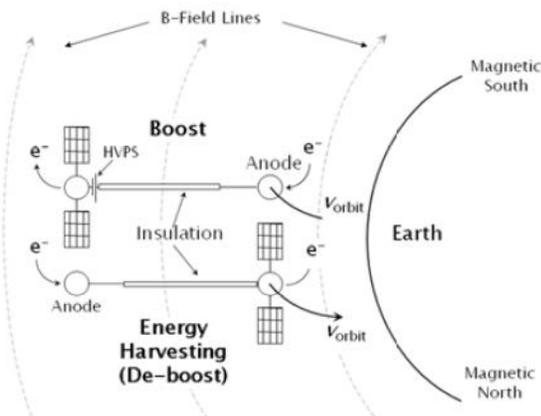

Fig. 3 Electrodynamic tether in energy harvesting and boosting modes.

## V.    Conclusion

Energy harvester had its very important place in space technologies long before they started to pave their way into consumer, automotive, industrial control and biomedical application. Even if very well described from theoretical point of view, very well simulated and with decades of heritage of successful exposure to most severe environmental conditions, space energy harvesters designs do not develop so quickly and do not evolve in different directions as their terrestrial counterparts. It seems that, once again, part of industrial and political activity which should be the avant-garde of technology, science and curiosity - space exploration is at the same time one of most conservative of all human activities. Space related conservatism is originating at reluctance to take risk, especially when it paired with high cost – of components, of qualified work force and of launch services.

At least dawn of nano-satellites (and smaller ones), bring hope that some of the innovative solution, especially tether based, will pave their way into everyday satellite operations, enabling more power generation and therefore more diverse and sophisticated spacecraft functions, and longer orbital life time.